\documentstyle[prb,twocolumn,aps]{revtex}

\def\UPt3{UPt$_3$}

\begin{document}
\draft
\twocolumn[\hsize\textwidth\columnwidth\hsize\csname@twocolumnfalse\endcsname

\title{Comment on ``Magnetic field effects on neutron diffraction in the 
antiferromagnetic phase of \UPt3''}

\author{B. F\aa k}
\address{ISIS Facility, Rutherford Appleton Laboratory, Chilton, 
Didcot, Oxon OX11 0QX, Great Britain}

\author{N. H. van Dijk}
\address{Delft University of Technology, Interfaculty Reactor 
Institute, Neutron Scattering and Mossbauer Spectroscopy Department, 
Mekelweg 15, 2629 JB Delft, The Netherlands}

\author{A. S. Wills\cite{ILL}} 
\address{Commissariat \`a l'Energie Atomique, D\'{e}partement de 
Recherche Fondamentale sur la Mati\`{e}re Condens\'{e}e, SPSMS/MDN, 
38054 Grenoble, France}

\date{5 June 2001}
\maketitle

\begin{abstract}
Moreno and Sauls have recently tried to re-analyze earlier neutron 
scattering studies of the antiferromagnetic order in \UPt3\ with a 
magnetic field applied in the basal plane.  In their calculation of 
the magnetic Bragg peak intensities, they perform an average over 
different magnetic structures belonging to distinct symmetry 
representations.  This is incorrect.  In addition, they have mistaken 
the magnetic field direction in one of the experiments, hence 
invalidating their conclusions concerning the experimental results.
\end{abstract}

\pacs{PACS number(s): 75.25.+z, 74.70.Tx, 75.30.Mb, 61.12.-q, 75.20.Hr}
]

Neutron elastic scattering measurements of the magnetic order in 
\UPt3\ have shown that a high magnetic field applied in the basal 
plane or along the hexagonal $c$-axis has virtually no effect on the 
size of the magnetic moment, the N\'eel temperature, or on the 
magnetic structure.\cite{Lussier,vanDijk} Within the precision of 
these measurements, no change of the domain 
populations\cite{Lussier,vanDijk} or of the moment direction in the 
basal plane\cite{Lussier} was observed.  Recently, Moreno and Sauls 
(MS)\cite{MS} have tried to re-analyze the two experiments in Refs.  
\onlinecite{Lussier} and \onlinecite{vanDijk} under the assumption 
that the pinning energy of the domain walls is larger than the 
in-plane anisotropy.  Unfortunately, the actual analysis of MS is 
incorrect for reasons that will be discussed below.

\UPt3\ orders antiferromagnetically below $T_N= 6$ K with a 
propagation vector {\bf k} = (0.5,~0,~0) and the Fourier component of 
the moment {\bf m}$_k$ parallel to {\bf k}.  For a sample without 
strain and in zero magnetic field, neutron scattering cannot 
distinguish between a single-k structure with three K domains and 
multi-k structures with or without domains.  Recent neutron scattering 
measurements under uniaxial pressure\cite{uniaxial} indicate that the 
magnetic structure is single-k, and we will only discuss this case in 
this Comment.  We restrict ourselves also to magnetic fields in the 
basal plane (Ref.\ \onlinecite{vanDijk} also treated $H||c$), as this 
is the only case analyzed by MS. Since the temperature dependence of 
the moment is smooth without any jump at the transition temperature 
and there is no evidence of any hysteresis or latent heat, we assume 
that the transition is second order.  We also assume that the moment 
is static and that the crystal structure in the paramagnetic phase is 
hexagonal with space group $P6_3/mmc$ ($D^4_{6h}$), although a lower 
trigonal symmetry was recently reported.\cite{NW} These assumptions 
were also made by MS.

Group theory analysis\cite{Bertaut,Izyumov-book,Wills-Sarah} indicates 
that for a propagation vector {\bf k} = (0.5,~0,~0) the magnetic 
representation that describes a magnetic moment at the U position 
(2{\it c}) can be decomponsed into 6 irreducible representations (IRs) 
of order one.  A ferromagnetic alignment of the moments within the 
unit cell (shown by neutron scattering measurements) is compatible 
with only 3 of these, namely $\Gamma_2$, $\Gamma_4$, and 
$\Gamma_6$.\cite{Kovalev} Application of the Landau theory for a 
second-order phase transition provides an important simplification to 
the analysis of the resulting magnetic structure because it requires 
that only one IR becomes critical.  Consequently, we can limit the 
{\it symmetry-allowed} magnetic structures to those defined by a 
single IR. As each of these IRs has only one basis vector associated 
with it in the present case, we find immediately that the moments are 
fixed along specific crystallographic directions.  The corresponding 
moment directions (assuming a single-k structure) are parallel to {\bf 
k}, perpendicular to {\bf k} in the basal plane, and parallel to the 
$c$ axis, respectively.  Since the {\bf Q} = (0.5,~0,~0) magnetic 
Bragg reflection is absent in neutron scattering measurements, the 
antiferromagnetic phase is described by $\Gamma_2$ with the moment 
parallel to the propagation vector.  In this case, there are no S 
domains,\cite{Sdomains} i.e. there is only one possible orientation of 
the moment {\bf m} with respect {\bf k}.  There are however three K 
domains, corresponding to the three equivalent orientations of {\bf k} 
in the basal plane: {\bf k}$_1$ = (0.5,~0,~0), {\bf k}$_2$ = 
(0,~0.5,~0), and {\bf k}$_3$ = (0.5,~-0.5,~0).  For unstrained 
(annealed) samples, the three K domains have equal population, as seen 
from the intensity in neutron scattering measurements.\cite{Hayden} 
This is the standard picture of the magnetic order in \UPt3\ as 
observed by neutron and x-ray scattering.

When a magnetic field is applied within the basal plane, the K domain 
with moments perpendicular to the applied field is favored over the 
other K domains.  For a sufficiently strong field, one would expect a 
repopulation of the different K domains.  Within current precision, 
this has not been observed by neutron scattering 
measurements.\cite{Lussier,vanDijk} However, recent measurements under 
uniaxial pressure suggest a domain repopulation.\cite{uniaxial}

Moreno and Sauls assume in their work that there are three ``domains'' 
for a given K domain, as illustrated in their Fig.~1.  However, the 
magnetic structures shown in Fig.~1b and c are not domains of the 
structure in Fig.~1a.  Rather, they are 2 S domains of a different 
magnetic structure.  While Fig.~1a, which corresponds to the actual 
magnetic structure of \UPt3, is described by the basis vectors 
associated with $\Gamma_2$, the structure shown in Fig.~1b and c 
corresponds to a mixing of those associated with $\Gamma_2$ and 
$\Gamma_4$.  Hence the structure presented as Fig.~1a has a different 
symmetry than that shown in Fig.~1b and c.  Since ordering under 
$\Gamma_2$ and $\Gamma_4$ involves more than one IR, the magnetic 
structure shown in Fig.~1b and c must involve a first-order 
transition, in contrast to the structure of Fig.~1a which is 
compatible with a second-order phase transition.  Although it appears 
as if the magnetic structures shown in Fig.~1 have the same energy, 
they have different symmetries.  Hence, one would expect that either 
the structure in Fig.~1a or the structure in Fig.~1b and c is 
established.  The absence of the {\bf Q} = (0.5,~0,~0) magnetic Bragg 
reflection in neutron scattering data shows unambigiously that the 
structure shown in Fig.~1a is established.  It is also the only one of 
the structures shown in Fig.~1 that is compatible with a second-order 
phase transition (provided that the non-magnetic space group is 
$P6_3/mmc$).  Even if the magnetic phase transition were first order 
so that the restrictions of Landau theory no longer apply, the 
structure of Fig.~1a has still a different symmetry from that in 
Fig.~1b and c.

In their actual analysis of the experimental data, MS evaluate the 
ratio $r$ of the magnetic Bragg peak intensities in field and in zero 
field, given by Eq.  (4) in Ref.~\onlinecite{MS}.  However, they 
average Eq.~(4) over the two {\it different structures} shown in 
Fig.~1.  This is clearly wrong.  Since the moment is parallel to the 
propagation vector (see Fig.~1a), there are no S 
domains,\cite{antiphase} and there should be no averaging.  If the 
moment were not parallel to {\bf k}, but still in the basal plane 
forming an angle $\alpha$ with respect to the propagation vector 
(which would require a first-order transition), the two S domains 
corresponding to $+\alpha$ and $-\alpha$ (these are illustrated in 
Fig.~1b and c for the case of $\alpha= 60^\circ$) should be averaged.  
The incorrect averaging over different magnetic structures in MS 
invalidates their analysis of both Ref.\ \onlinecite{Lussier} and 
\onlinecite{vanDijk}.  In particular, the results given in Eqs.~(5-8) 
are all incorrect.

A second problem is that MS's analysis of the work by van Dijk et 
al.\cite{vanDijk} assumes wrongly that the applied field was along the 
particular $a$ axis that was at 30$^\circ$ with respect to the 
observed moment.  However, van Dijk et al.  clearly stated that they 
measured the {\bf Q} = (0.5,~0,~1) magnetic Bragg peak using a 
vertical-field magnet, which means that the magnetic field was applied 
perpendicular to the horizontal scattering plane, and hence at 
90$^\circ$ with respect to the moments of the studied Bragg peak.  The 
field was thus along the $a$ axis that in the notation of Ref.\ 
\onlinecite{vanDijk} can be labeled (0,~1,~0) in real space or 
(-1,~2,~0) in reciprocal space.  In this geometry, there is no reason 
for the moment in the {\bf k} = (0.5,~0,~0) domain to rotate, so no 
intensity change is expected.  However, if the other K domains would 
be depopulated and instead contribute to the {\bf k} = (0.5,~0,~0) 
domain, the intensity at {\bf Q} = (0.5,~0,~1) would increase by a 
factor of three (for a full domain repopulation) as stated by van Dijk 
et al.

In the experiment by Lussier et al.,\cite{Lussier} the lower magnitude 
of the applied field allowed the use of a horizontal field magnet, 
which gives a much larger choice of geometries.  Their analysis, which 
is correct, shows that for a field of 3.2 T in the basal plane there 
is no domain repopulation, not even for a field-cooled sample where 
the pinning energy is irrelevant, as only the most favored domain will 
form on cooling through the N\'eel temperature.  Notably, they also 
did not observe any moment rotation.  

MS also suggest that it is not known whether the Fourier component of 
the magnetic moment is parallel to the propagation vector in zero 
field.  However, Hayden et al.\cite{Hayden} showed beyond any doubt 
that {\bf m}$_k$ is parallel to {\bf k}.  For the same sample, they 
first showed that the K domains are equally populated, by measuring at 
three different Bragg peaks, each corresponding to a different {\bf k} 
vector.  Next, they showed that a magnetic Bragg peak with {\bf Q} 
$\parallel$ {\bf k} in one of these domains has zero intensity, which 
proves unambiguously that {\bf m}$_k$ is parallel to {\bf k}.  The 
same result has been found by other groups, including ours.

In summary, although the scenario with a large pinning energy and the 
discussion of the symmetry breaking properties of a triple-k structure 
are interesting, the actual analysis by Moreno and Sauls of the field 
dependence of the magnetic Bragg peak intensities in \UPt3\ is 
incorrect, as they perform an average over different magnetic 
structures that are symmetry inequivalent.  There is no experimental 
evidence that the moment rotates away from the propagation vector when 
a magnetic field is applied in the basal plane.  Also, such a rotation 
is not compatible with the symmetry properties of a second-order 
phase transition.

We have benefited from discussions with F. Bourdarot and J. Schweizer 
on S and K domains.


\begin{references}
\bibitem{Sdomains}
S domains, also termed orientational domains, differ only in the 
choice of the equivalent local axes.  They are necessarily of the same 
symmetry and so may be described by the same irreducible 
representation.
\bibitem{antiphase}
There are also two so-called 180$^\circ$ domains, which differ only by 
the phase of the stacking sequence (i.e. $+-+-$ {\it vs} $-+-+$), but 
since they are indistinguishable, we do not count them here.
\end{references}
\end{document}